\documentclass[a4paper,UKenglish]{lipics-v2019}
\pdfoutput=1
\usepackage[dvipsnames]{xcolor}

\usepackage{amsmath,amssymb,amsthm,graphicx,fixmath,tikz,latexsym,xspace}

\newtheorem{stylised}[theorem]{Hypothetical stylised fact}

\title{Data-Driven Classifications of Video Game Vocabulary}

\titlerunning{Data-Driven Classifications of Video Game Vocabulary} %TODO optional, please use if title is longer than one line

\author{Nicolas Grelier}{Focus Entertainment, France}{nicolas.grelier@focusent.com}{}{}
\author{Stéphane Kaufmann}{Focus Entertainment, France}{stephane.kaufmann@focusent.com}{}{}

\authorrunning{N. Grelier, S. Kaufmann} %TODO mandatory. First: Use abbreviated first/middle names. Second (only in severe cases): Use first author plus 'et al.'

\Copyright{Nicolas Grelier, Stéphane Kaufmann} %TODO mandatory, please use full first names. LIPIcs license is "CC-BY";  http://creativecommons.org/licenses/by/3.0/

\ccsdesc[500]{Applied computing~Publishing}

\keywords{taxonomy, meronomy, video game genres, Steam tags, game analytics}

\category{} %optional, e.g. invited paper

\relatedversion{} %optional, e.g. full version hosted on arXiv, HAL, or other respository/website
\supplement{}%optional, e.g. related research data, source code, ... hosted on a repository like zenodo, figshare, GitHub, ...

\nolinenumbers %uncomment to disable line numbering

\hideLIPIcs  %uncomment to remove references to LIPIcs series (logo, DOI, ...), e.g. when preparing a pre-final version to be uploaded to arXiv or another public repository

\EventEditors{John Q. Open and Joan R. Access}
\EventNoEds{2}
\EventLongTitle{42nd Conference on Very Important Topics (CVIT 2016)}
\EventShortTitle{CVIT 2016}
\EventAcronym{CVIT}
\EventYear{2016}
\EventDate{December 24--27, 2016}
\EventLocation{Little Whinging, United Kingdom}
\EventLogo{}
\SeriesVolume{42}
\ArticleNo{23}
\begin{document}

\maketitle

\begin{abstract}
As a novel and fast-changing field, the video game industry does not have a fixed and well-defined vocabulary. In particular, game genres are of interest: No two experts seem to agree on what they are and how they relate to each other. We use the user-generated tags of the video game digital distribution service Steam to better understand how players think about games. We investigate what they consider to be genres, what comes first to their minds when describing a game, and more generally what words do they use and how those words relate to each other. Our method is data-driven as we consider for each game on Steam how many players assigned each tag to it. We introduce a new metric, the priority of a Steam tag, that we find interesting in itself. This allows us to create taxonomies and meronomies of some of the Steam tags. In particular, in addition to providing a list of game genres, we distinguish what tags are essential or not for describing games according to players. Furthermore, we provide a small group of tags that summarise all information contained in the Steam tags.
\end{abstract}

\section{Introduction}

In the last two decades, game analytics has gained a lot of interest~\cite{su2021comprehensive}. According to Su, Backlund and Engström, ``game analytics is the process of identifying and communicating meaningful patterns that can be used for game decision making''~\cite{su2021comprehensive}. They present a classification of game analytics into four groups: game development analytics, game publishing analytics, game distribution channel analytics and game player analytics. In this paper, we focus on game player analytics. Our aim is to understand the vocabulary used by players to speak about video games. More precisely, we are interested in how players describe games: what are the most important words, what are the most used words, and so on.

In our study, we rely exclusively on Steam tags: groups of words that players can freely assign to games. Some examples are \emph{Strategy}, \emph{Card Battler}, \emph{Agriculture}, \emph{Ninja}, \emph{Difficult} and \emph{Free to Play}. In total, there exist $427$ Steam tags. For a more comprehensive explanation of Steam tags, please refer to Subsection~\ref{subsec:pres_data}.

\subsection{From beliefs to facts}
In~\cite{drachen2016stylized}, Drachen {\em et al.} observe that game analytics ``is in its infancy and the available knowledge is heavily fragmented, not the least due to Game Analytics interdisciplinary nature and the lack of knowledge sharing between academia and industry''. As industry researchers, the goal of our paper is to build bridges with academia. Drachen {\em et al.} add that ``the root cause of these problems is the lack of a framework for organizing current knowledge and prioritising interesting problems''~\cite{drachen2016stylized}. They introduce the concept of \emph{stylised facts} for mobile game analytics (although we believe that the concept is essential to game analytics in general) in order to better connect industry and academia.

However, they argue that the available knowledge and data are too fragmented and incomplete to establish stylised fact. Thus, they introduce two ``proto-stylised fact concepts which describe situations where less empirical validation is available: \emph{beliefs} and \emph{hypothetical stylised facts}''~\cite{drachen2016stylized}. The aim is to ``provide
a roadmap for structuring current knowledge and building towards a situation where stylised facts can be generated and validated''. Beliefs are statements supported by virtually no empirical evidence, whereas hypothetical stylised facts are supported by some empirical evidence. However, they are still not stylised facts, since ``while there maybe some available empirical evidence supporting these hypothetical stylised facts, it is clearly not enough to rigorously, generally, support them''. Establishing hypothetical stylised facts help in obtaining stylised fact, which is essential for the research field (see~\cite{drachen2016stylized} for a detailed argumentation).

In this paper, we provide empirical evidence for several hypothetical stylised facts. Some of which we introduce, while others were already beliefs in the industry, for instance that genre is one of the most crucial pieces of information when describing a video game. We aim at presenting our findings in a quantitative manner, and to provide an interpretation of the metrics we introduce.

\subsection{Taxonomies and meronomies}

In Section~\ref{sec:taxonomy}, we present a data-driven taxonomy of all Steam tags. This consists in doing a hierarchical classification, where we divide the tags into categories and subcategories. Those are referred to as ``taxa''. Each tag is put into one of the taxa of lower rank, which are themselves grouped into taxa of higher rank. Our taxonomy has a tree-like structure, but others may be more network-like, with one child having several parents.

We also provide in Section~\ref{sec:meronomy} two data-driven meronomies of some Steam tags. A meronomy differs from a taxonomy as it deals with part-whole relationships, whereas a taxonomy is a classification of discrete elements. Using mathematical notation, a meronomy can be thought of as a partially ordered set where the elements are sets and the relation considered is inclusion (we have $A \leq B$ if $A$ is a subset of $B$). Here, our aim is to better understand how some tags relate to each other. Using a data-driven approach, we establish some relation between tags, for instance that \emph{2D Platformer} is a subtag of \emph{Platformer}, and that \emph{Sailing} is a subtag of \emph{Naval}.

\subsection{A focus on game genres}

A part of our study concerns game genres. In particular, we are interested in finding a complete list of game genres, and to understand how much genre matters to players. Arsenault observes that many websites allow the user to search through video games while filtering by genre~\cite{arsenault2009video}. However, no two websites seem to agree on what the list of genres is. For instance, he mentions GameSpot's system (at the time of his writing) that does a taxonomy of the genres: genres are subdivided into subgenres, thus creating $157$ categories. He gives some examples:
\begin{itemize}
    \item \emph{Action} $\rightarrow$ \emph{Shooter} $\rightarrow$ \emph{First-Person} $\rightarrow$ \emph{Fantasy},
    \item \emph{Miscellaneous} $\rightarrow$ \emph{Puzzle} $\rightarrow$ \emph{Action},
    \item \emph{Strategy} $\rightarrow$ \emph{Real-Time} $\rightarrow$ \emph{Fantasy}.
\end{itemize}

Arsenault notes that ``while every category is unique and independent in principle, some of the lower-tiered
descriptors appear as sub-branches of multiple genres''~\cite{arsenault2009video}. He adds that ``the levels or branches themselves are not named, which means there is no basis on which to compare them'', and ``comparing all 2nd-level classes would probably amount to comparing apples to oranges to shoes to faith''. He concludes: ``One thing that these different taxonomies highlight is the fluidity and impreciseness of the concept of genre itself, and how it is used in actually describing games''. The same issues are noted by Heintz and Law~\cite{heintz2015game}:

\begin{itemize}
    \item Genres are not clearly or consistently defined,
    \item The relation between genres is unknown,
    \item Definitions are based on completely different aspects,
    \item Different sources use different sets of genres.
\end{itemize}

%%%TO BE WRITTEN: more detailed list of who presented a list/taxonomy of genres.

Steam provides a set of tools for developers and publishers named Steamworks. They provide a taxonomy of Steam tags~\cite{steamworks2021steam}. Tags are divided into several taxa: Genres, Visual Properties, Themes \& Moods, Features, Players, Other Tags, Software, Assessments, Ratings etc, Hardware/Input and Funding etc. The taxon Genres, which is our main focus here, is split into three taxa: Super-Genre, Genre and Sub-Genre. This taxonomy is to be understood with the broad definition of a taxonomy, in which a child may have several parents. For instance, \emph{Heist} is a sub-genre and also a theme. \emph{Flight} is a sub-genre and also a feature. \emph{Experimental} is a super-genre and a genre. One objective of this paper is to challenge Steam's taxonomy and investigate whether players concur with it.

Classifying video games into genres is valuable for studying and comparing meaningful categories. For instance, in the literature we can find studies on what genres are the most successful~\cite{foxman2020virtual, qaffas2020operational} (and also~\cite{samarasinghe2021data} although the study is about board games instead of video games), how considering games from the same genre is important for deciding on a release date~\cite{engelstatter2018strategic,grewal2022empirical}, how genre relates to players motivations, gender and localisation~\cite{lucas2004sex,ratan2021gender,tekofsky2016effect, vermeulen2016play}, or what are their usability profiles~\cite{pinelle2008using}. All these studies use a list of genres, but do not discuss much (or not at all) whether it makes sense to discuss about genres, whether the list is comprehensive, and so on.

A more industry-oriented reason for establishing a comprehensive list of game genres, along with definitions and relationships between the genres, comes from trailer making. Lieu, who was involved in the creation of the trailers of \emph{Half-Life: Alyx}, \emph{Among Us} and \emph{The Long Dark} among others, asserts that a trailer of a video game must first clearly establish its genre~\cite{lieu2021game}. Therefore, it is crucial to have an exact understanding of what a genre is. Moreover, Lieu details what a trailer should show according to the game genre~\cite{lieu2021trailer}. Similarly, Carless, who is the founder of GameDiscoverCo, a video game discoverability consultancy firm, believes that game genre is ``what [their] users want to see information about''. In a database meant for people working in the video game industry, Carless ``care[s] most that the main genres are in \emph{Genre} and are easy to see and sort'' ([Carless, personal communication]). 

\subsection{The Steam tags}
\label{subsec:pres_data}
In this paper, we rely solely on Steam data. More precisely, we study the user-generated tags on Steam. There exists a total of $427$ tags that players have assigned to the $50757$ games in Steam. Those tags, consisting of groups of words or acronyms like \emph{Action RPG}, \emph{MOBA} or \emph{Open World}, are freely assigned to games by players, who can also invent new ones. 

There exist many papers relying on Steam data, for instance Steam players reviews~\cite{lin2018empirical,petrosino2022panorama,pirker2022virtual,sobkowicz2016steam,zuo2018sentiment} or Steam user accounts~\cite{baumann2018hardcore,becker2012analysis,o2016condensing,sifa2015large}. Some of these studies use Steam tags, but there are only used as tools for filtering games, and are not the topic of study. To the best of our knowledge, there are only three papers that truly study Steam tags per se. First, Windleharth {\em et al.} do in~\cite{windleharth2016full} a conceptual analysis of Steam tags. They discuss and classify Steam tags into categories, some containing many elements like Gameplay genre, and some containing only one elements like Relationships which contains only the tag \emph{Remake}. Windleharth {\em et al.} rely on their own expertise to determine what tags are genre tags. In this paper, we use a data-driven approach with the same goal in mind. A second paper that studies Steam tags is~\cite{li2020preliminary}, in which Li and Zhang examine how genre tags are related to each other. Finally, in~\cite{li2020towards}, Li uses correlation between Steam tags related to gameplay to obtain a list of $29$ ``factors''. Those factors are not necessarily Steam tags but rather groups of highly correlated Steam tags. For instance, the factor \emph{Rogue} contains the Steam tags \emph{Dungeon Crawler}, \emph{Rogue-like} and \emph{Rogue-lite}. We compare our method to theirs in Section~\ref{sec:priority}, and argue why ours is more reliable. We also compare our results with theirs in Sections~\ref{sec:taxonomy} and~\ref{sec:meronomy}.

All players are free to assign to games some of the already-existing Steam tags, or even to add new ones. For a game, Steam shows ``tags that were applied to the product by the most users''. Steam shows at most $20$ tags for any game, sorted by decreasing order of number of players who assigned the tag. The exact number of players who assigned a tag is not directly shown on Steam, but can be found in the source code of the page. To obtain the data of how many players assigned a tag to a game, we used SteamSpy's API. There are $13848$ games with $20$ tags shown on Steam, and therefore $50757-13848=36909$ games with fewer than $20$ tags. One can make the experience of adding a new tag to a game with fewer than $20$ tags, and have the surprise of not having their newly added tag shown in Steam. The reason for that is that Steam only shows tags that were assigned by at least $5$ players. Steam does not explicitly mention this, but we could check it by going over all games. Since anyone can add any set of nonsensical words as a tag, this is probably for the best.  

It is important to note that Steam does not offer a definition of the tags, one sufficient reason being that tags are freely added and invented by players (as can been seen in the inconsistent capitalisation and hyphenation of the tags). Therefore, players interpret by themselves the meanings of the tags. For our study, this comes with pros and cons. Since ``genres are not clearly or consistently defined''~\cite{heintz2015game}, we may, thanks to Steam tags, find consensuses about what a given genre is. We believe that there exists some common understanding of what, say, an adventure game is. By an equivalent of the law of large numbers, Steam tags can help decipher what the definition is. However, there are also cases when players are clearly referring to distinct things while using the same tag, which perturbs our analysis. One easily spotted example is the tag \emph{Football}, which sometimes refers to soccer, and sometimes to American football. An another example is the tag \emph{Fighting}. For some games, it seems that the tag is used to state the genre of the games, whereas in other games it may be used to merely inform that there is some fighting in the game. This motivates why some tags have several parents in Steamworks' taxonomy~\cite{steamworks2021steam}.

Several of our results rely on computing the Pearson correlation coefficient between tags. We point out another difficulty we face, that stems from the nature of the database. One would expect some pairs of tags to have an extremely low correlation (around $-1$), for instance: \emph{2D} and \emph{3D}, \emph{First-Person} and \emph{Third Person}\footnote{We follow Steam spelling with a hyphen in \emph{First-Person} but none in \emph{Third Person}.}, \emph{Singleplayer} and \emph{Multiplayer}, and so on. Indeed, a priori, a game is either in 2D or 3D, either has a first-person or third person viewpoint, and is either a singleplayer or a multiplayer game. To compute the correlation, we create a matrix with one entry per game, and one column per tag. There is a value of $1$ if the tag is assigned to the corresponding game, and $0$ otherwise. We compute a correlation of $-0.17$ for \emph{2D} and \emph{3D}, $0.04$ for \emph{First-Person} and \emph{Third Person}, and $0.10$ for \emph{Singleplayer} and \emph{Multiplayer}. We investigated what was the cause of these surprisingly high correlation coefficients. There are games that use 2.5D or isometric representations. Those are mixes of 2D and 3D features. We found that generally, whenever the tag \emph{2.5D} or \emph{Isometric} is assigned to a game, the tags \emph{2D} and \emph{3D} are also (wrongly) assigned to the game by players. This is why the correlation is closer to $0$ than to $-1$. Concerning the viewpoint, there are many shooter games using the third person view that change for a first-person view when the player is aiming. In those games, we found that both tags \emph{First-Person} and \emph{Third Person} are assigned. Similarly, if a game contains a singleplayer campaign and a multiplayer mode, then both tags \emph{Singleplayer} and \emph{Multiplayer} are usually assigned. In conclusion, one has to be very careful when exploiting pairs of tags with low correlation, since one might miss a few relevant pairs.

Steam does not provide the information of how many players assigned tags to a given game. However, the number of players who assigned the most given tag of a game $G$ gives us a lower bound on the number of players who assigned tags to $G$. Using this method, we know that there are at least $15843$ games for which at least $100$ players assigned some tags.

We said that Steam shows at most $20$ tags for a game. This is not exactly true in the source code. Indeed, there are a few games, $235$ to be exact, with $21$ tags. They all share a common property: The $21$st tag is \emph{VR Only}, and is assigned by precisely one player. This is probably a feature added by Steam. We make the choice of removing this tag from the database, since it was not added by real players.

\subsection{Our contributions}

In Section~\ref{sec:priority} we present a new concept: the priority of a tag for a game. We show that this is a useful notion, for we establish as a hypothetical stylised fact that tags with the highest priority are what players think about first when describing the game (see Section~\ref{sec:priority} for a formal definition of priority and a formal statement of the hypothetical stylised fact). On the other hand, tags with low priority are what players deem interesting about the game, but which only come later to their minds. Using this concept of priority, we give empirical evidence in Section~\ref{sec:taxonomy} for the following hypothetical stylised fact:

\begin{stylised}\label{fact:genre}
Game genre is what players think about first when describing a game.
\end{stylised}

Consequently, we believe that this is also what they want to hear about first when discovering a new game, confirming the soundness of Lieu's practices in trailer making~\cite{lieu2021game}. On a related note, using a data-driven approach we establish a list, as comprehensive as possible, of game genres. Our aim here is to encompass all words that players may consider as game genres. The list can be found in Appendix~\ref{app:list_genres}.

We provide in Section~\ref{sec:meronomy} a meronomy of the tags (not necessarily about genre) with broadest meanings. In particular, we find a set of seven tags with broadest meanings, that we name the ``capital tags''. There are four properties that we want the capital tags to satisfy. First, they should form a rather short list. As we have $7$ capital tags out of the $427$ Steam tags, this is clearly satisfied. Secondly, for any game, at least one of these tags should be assignable to the game. Thirdly, those tags should not be redundant. Finally, any other tag should be a more refined version of some of the capital tags. We show in Section~\ref{sec:meronomy} that the capital tags satisfy this three last properties. This implies that the capital tags summarise all information contained in Steam tags.

We provide a second meronomy in Section~\ref{sec:meronomy}, that aims at reducing the number of Steam tags without loosing information, by merging synonymous tags together. We propose a method for doing so as well as finding a representing tag for each group of merged tags.

\section{The priority of Steam tags}
\label{sec:priority}

In~\cite{li2020preliminary}, Li and Zhang defines a graph that aims to accurately represent the Steam tags. There is one vertex per tag, and two tags are connected by an edge if they are both assigned to a same game. The edges are weighted according to how many games have both tags assigned to them. The authors discard some edges by using the ``Narrow by tag'' feature in Steam. We find the authors' results difficult to interpret due to their chosen methodology. First, it is not clear what the Steam ``Narrow by tag'' feature does. Additionally, the authors connect two tags $T$ and $T'$ with an edge even if for the game $G$ to which they are both assigned, the tag $T$ is the most assigned tag to $G$ and $T'$ is the least assigned tag to $G$, without discussing the rationale behind this decision. It seems wrong to us to connect the two tags with an edge since $T$ is apparently essential for describing $G$ whereas $T'$ is not so much relevant.

In~\cite{li2020towards}, Li computes correlation between tags. The analysis is more precise than in~\cite{li2020preliminary} since when a tag $T$ is assigned to a game, a weight is associated to $T$. If for a given game $G$, the tag $T$ is the tag most assigned to $G$, then $T$ obtains a weight of $20$. If $G$ has $20$ tags and $T$ is the least assigned, it obtains a weight of $1$. More generally, the $n$-th most assigned tag obtains a weight of $20-n+1$. We point out two issues. First, Li does not discuss whether it makes sense to use these weights, and how to interpret them. Secondly, the distributions of the number of players who assigned tag varies greatly over all games. For instance, the two most assigned tags to the game \emph{HITMAN 2} are \emph{Stealth} ($10373$ players) and \emph{Assassin} ($819$ players). The ratios of the number of players for these tags is therefore $819/10373\approx 0.08$. If we do the same with the same with the game \emph{Golf It!}, we find that the most assigned tag is \emph{Multiplayer} ($249$ players) and the second most assigned is \emph{Mini Golf} ($243$ players). This time, the ratio is $243/249\approx 0.98$. With Li's method, those ratios would be $19/20 = 0.95$ for both games~\cite{li2020towards}.

In this section, we define, interpret and study a new metric, that will be used in all later sections for the taxonomies and meronomies. It is a more refined version of what was proposed by Li~\cite{li2020towards}. For each Steam tag, we define what we call its \emph{priority} for a given game.

\begin{table}
\centering
\begin{tabular}{|l|l|l|}
\hline
tag         & number of players & priority \\ \hline
\emph{Adventure}    & 1000              & 1        \\
\emph{Puzzle}      & 750               & 0.75     \\
\emph{2D}          & 500               & 0.5      \\
\emph{Atmospheric} & 100               & 0.1      \\
\emph{3D}          & 0                 & 0        \\ \hline
\end{tabular}
\caption{A fictional example of tags assigned to a game with their priorities}
\label{tab:priority}
\end{table}

\begin{definition}\normalfont
The priority of a tag $T$ for a given game $G$ is a score that ranges from $0$ to $1$. Let $t_T$ denote the number of players who assigned to the game $G$ the tag $T$. Similarly, let $t_{\max}$ denote the maximum over all tags assigned to $G$ of the number of players who assigned the tag. The priority of $T$ for $G$ is equal to $t_T/t_{\max}$. A fictional example is shown in Table~\ref{tab:priority}.
\end{definition}

In particular, any game with at least one tag has a tag with priority $1$: the tag that was assigned to it by the largest number of players. If a tag was not among the $20$ most assigned tags to a given game, then its priority is $0$ (recall that Steam only shows the $20$ most assigned tags). We choose this metric as we are interested in identifying how players think about games. Let us consider all positive priorities over all games and tags. We compute a mean of $0.60$, a median of $0.61$ and a standard deviation of $0.28$. This shows that the priority contains some information, as the standard deviation is quite high. There are two compatible reasons as to why not all tags have the same priority for a given game $G$. First, it may be that some players assign much fewer tags than other players. This would imply that the tags with highest priority correspond to what players think about first when describing $G$. The second reason would be that players do not perfectly agree on what tags are best suited to describe $G$. This second reason would imply that the tags with highest priority are what players agree upon for describing $G$, and that their opinions diverge on the tags with low (but non-zero) priority. We argue that the first reason is the most important, and present the following hypothetical stylised fact:

\begin{stylised}\label{fact:priority}
For a given game $G$, the tags with highest priority correspond to what players think about first when describing~$G$.
\end{stylised}

We provide empirical evidence supporting Hypothetical Stylised Fact~\ref{fact:priority}. The fact is hypothetical as defined in~\cite{drachen2016stylized} because our methods for supporting it cannot establish it for good. This could be done if we had access to a database of the tags assignment per player, but this is not publicly available.

\begin{figure}[htb]     
\centering
\begin{subfigure}{\linewidth}
  \centering
  \includegraphics[scale=0.5]{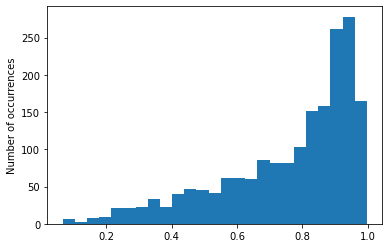}
  \caption{Histogram of the ratios of the priority of \emph{FPS} and \emph{Shooter} for the $1867$ games whose \emph{FPS} priority is higher than the one of \emph{Shooter}.}
  \label{fig:FPS_priority} 
\end{subfigure}
\begin{subfigure}{\linewidth}
  \centering
  \includegraphics[scale=0.5]{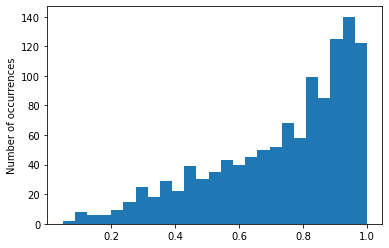}
  \caption{Histogram of the ratios of the priority of \emph{Online Co-Op} and \emph{Co-op} for the $1171$ games whose \emph{Online Co-Op} priority is higher than the one of \emph{Co-op}.}
  \label{fig:coop_priority} 
\end{subfigure}
\caption{Some empirical evidence for Hypothetical Stylised Fact~\ref{fact:priority}.}
\label{fig:fact_priority}
\end{figure}

We consider two tags: \emph{Shooter} and \emph{FPS}. ``FPS'' stands for ``first-person shooter'', and is a subgenre of shooter game~\cite{adams2006fundamentals,voorhees2014routledge}. Therefore, anyone who thinks that a given game $G$ is an FPS would also agree that it is a shooter game. We consider all of the $1867$ games to which both tags \emph{Shooter} and \emph{FPS} are assigned, with \emph{FPS} having the highest number of players. If Hypothetical Stylised Fact~\ref{fact:priority} were wrong, for each game we should have very similar priorities for both tags. Indeed all players who assigned the tag \emph{FPS} would also assign the tag \emph{Shooter}. In contrast, we believe that some players assign only a few tags, corresponding to what comes first to their mind. Under this hypothesis, we would expect some players to only assign the tag \emph{FPS}, but not the tag \emph{Shooter} as it is less important to them, or because they consider it to be already implied by the tag \emph{FPS}. The histogram in Figure~\ref{fig:FPS_priority} illustrates the ratios between the priority of \emph{FPS} and \emph{Shooter}. We observe a long tail phenomenon; also the mean is equal to $0.76$ and the median to $0.83$. This implies that for half of the considered games, less than $83\%$ of those players who assigned the tag \emph{FPS} also assigned the tag \emph{Shooter}.

We apply the same method for the tags \emph{Online Co-Op} and \emph{Co-op}~\footnote{Steam tags are case sensitive. We follow the Steam capitalisation which may differ from tag to tag.}. Obviously, a game with an online co-op mode thereby contains a co-op mode. However, when we consider the $1171$ games to which both tags \emph{Online Co-Op} and \emph{Co-op} are assigned, with \emph{Online Co-Op} having the highest number of players, we observe again a long tail phenomenon in Figure~\ref{fig:coop_priority}. The mean is equal to $0.73$ and the median $0.80$. The natural explanation to this long tail phenomenon and to these low means and medians is Hypothetical Stylised Fact~\ref{fact:priority}: Many players only assign a few tags to a game, the ones that come first to their mind. We have shown here the results for only two pairs of tags, for which the numbers of corresponding games were important ($1867$ and $1171$, respectively). We tested other pairs, and for all of them we obtained similar results, but due to lack of space we do not show them here. Also, the number of games was significantly lower. Some examples of those pairs are \emph{Top-Down Shooter} and \emph{Top-Down} ($655$ games), \emph{Football} and \emph{Sports} ($165$ games), \emph{Traditional Roguelike} and \emph{Rogue-like}~\footnote{Once again we follow the spelling of Steam, where rogue-like is not always written with a hyphen.} ($102$ games).

\begin{figure}
    \centering
    \includegraphics[scale=0.5]{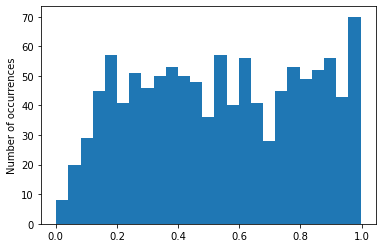}
    \caption{Histogram of the ratios of the priority of \emph{RPG} and \emph{Multiplayer} for the $1124$ games whose \emph{RPG} priority is higher than the one of \emph{Multiplayer}.}
    \label{fig:rpg_multi}
\end{figure}

In Figure~\ref{fig:fact_priority}, we observed histograms with a peak around a priority of $0.9$ and a long tail. Recall that we were considering pairs of tags where the first tag implies the second. We argued that causality was not perfectly reflected in the tag distributions, because of the long tail phenomenon. However, the position of the peak still indicates that there is a high correlation. As a comparison, let us do the same for two tags which are a priori not correlated. In Figure~\ref{fig:rpg_multi}, we applied the same method as before on the tags \emph{RPG} and \emph{Multiplayer}. Those are intuitively not correlated since there are multiplayer RPGs (for instance MMORPGs), but also many singleplayer RPGs. We observe that the histogram is rather constant; there is no strong peak around $1$. This is a first hint that useful information might be contained in the correlation between tag priorities. We use this idea of considering the correlation for our meronomies in Section~\ref{sec:meronomy}.

We point out an issue that we discovered concerning the priority. Although we still believe the notion to be of interest, and that it brings useful information, we think that the notion could be refined and improved in future work. It seems that for some reason unknown to us, the priority of some tags behave erratically. Let us consider for instance the tag \emph{Free to Play}. We consider three free to play games: \emph{Dota 2}, \emph{Team Fortress 2} and \emph{Counter-Strike: Global Offensive}. For the first two games, \emph{Free to Play} has a priority of $1$. However, the second tag with highest priority for \emph{Dota 2} is \emph{MOBA} with a priority of $0.33$, whereas for \emph{Team Fortress 2} it is \emph{Hero Shooter} with a priority of $0.99$. It is quite intriguing to us how the priorities of the second tags with highest priorities can be so different. This phenomenon occur regularly with the tag \emph{Free to Play}. We do not understand why the priority of the second most assigned tag can differ so much, when both games are equally free to play. We fear that for \emph{Dota 2}, the tag \emph{Free to Play} pushes all other priorities to very low values, thereby introducing noise in the data. Concerning \emph{Counter-Strike: Global Offensive}, the priority of \emph{Free to Play} is $0$. The reason for this can be easily understood: The game was released in $2012$ and made free to play only in $2018$. We believe that the tags were mostly assigned soon after the release of the game, and that is why \emph{Free to Play} is not amongst the $20$ most assigned tags. This motivates even further the fact that the method should be refined, with a particular emphasis on the tag \emph{Free to Play}.

\section{A taxonomy of Steam tags}
\label{sec:taxonomy}

In this section, we categorise all of the Steam tags through a taxonomy. In Section~\ref{sec:priority}, we defined the notion of priority of a tag for a game, and established that it contained some information. In particular, we gave empirical evidence for Hypothetical Stylised Fact~\ref{fact:priority}, stating that for a given game $G$, tags with high priorities correspond to what players think about first when describing $G$. In this section, we consider tags one by one, and for each tag we look at its priority distribution over all games for which the priority is non-zero. This allows us to define the taxa of higher rank into which we classify the Steam tags.

\subsection{The taxa of higher rank}
\label{subsec:taxonomy_high}

At the highest level, we have three taxa: High priority tags, Medium priority tags and Low priority tags. Let us start with the taxon Low priority tags.

\subsubsection{Low priority tags}

According to Hypothetical Stylised Fact~\ref{fact:priority}, those tags correspond to what few players deem interesting for describing a game, and that is not what they think about first. We define this taxon as the set of tags whose priority median is at most $0.45$. There are $80$ tags in this taxon. We chose this threshold using our own expertise of the video game industry. Tags with higher priority medians seem to us to come significantly faster in players' minds when describing a game.

\begin{table}[]
\centering
\scalebox{0.9}{
\begin{tabular}{|l|l|l|}
\hline
tag          & priority median & number of games \\ \hline
\emph{Masterpiece}  & 0.18            & 6               \\ \hline
\emph{Epic}         & 0.20            & 106             \\ \hline
\emph{TrackIR}      & 0.20            & 31              \\ \hline
\emph{Vikings}      & 0.22            & 45              \\ \hline
\emph{Reboot}       & 0.23            & 11              \\ \hline
\emph{Mod}          & 0.24            & 62              \\ \hline
\emph{Addictive}    & 0.24            & 409             \\ \hline
\emph{Cult Classic} & 0.25            & 305             \\ \hline
\emph{Kickstarter}  & 0.26            & 181             \\ \hline
\end{tabular}
}
\caption{Tags with the lowest priority medians.}
\label{tab:lowPrio}
\end{table}

\begin{table}[]
\centering
\scalebox{0.9}{
\begin{tabular}{|l|l|l|}
\hline
tag              & priority median & number of games \\ \hline
\emph{Singleplayer}     & 0.36            & 23376           \\ \hline
\emph{Multiplayer}      & 0.42            & 6371            \\ \hline
\emph{Retro}            & 0.43            & 4748            \\ \hline
\emph{VR}               & 0.44            & 4696            \\ \hline
\emph{Difficult}        & 0.39            & 4404            \\ \hline
\emph{Great Soundtrack} & 0.31            & 4245            \\ \hline
\emph{Co-op}            & 0.39            & 3153            \\ \hline
\emph{Controller}       & 0.40            & 2469            \\ \hline
\emph{Combat}           & 0.40            & 2135            \\ \hline
\end{tabular}
}
\caption{Low priority tags that are most assigned to games.}
\label{tab:lowPrio_highGames}
\end{table}

We present in Table~\ref{tab:lowPrio} the tags with the lowest priority median, along with the number of games to which they are assigned. One could argue that we do not have enough data to deal with tags that are assigned to very few games. However, for the taxon of Low priority tags, we think that if a tag was seldomly assigned, it supports even more the claim that this tag is not what players think about first when describing a game. Nonetheless, among the Low priority tags, there are some that are assigned to thousands of games. We show them in Table~\ref{tab:lowPrio_highGames}. We find those tags especially interesting. Players do not use those tags as their primary descriptors for a game, yet they are still assigned to many games. This gives us an interesting learning: Having a low priority does not mean that the tag is not important. Indeed, it is important for players to know whether a game has singleplayer or multiplayer modes. It is just not what is the most essential to describe a game. In Subsection~\ref{subsec:capital}, we delve deeper into the tags \emph{Singleplayer} and \emph{Multiplayer}.

\subsubsection{High priority tags}

\begin{figure*}
\centering
\begin{minipage}[b]{0.3\textwidth}
\centering
\includegraphics[scale=0.3]{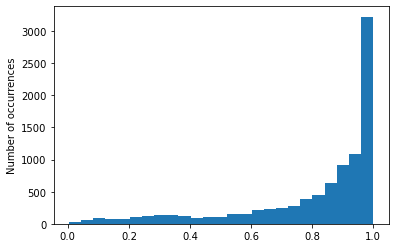}\\
\end{minipage}%
\begin{minipage}[b]{0.3\textwidth}
\centering
\includegraphics[scale=0.3]{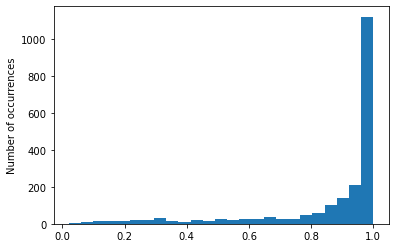}\\
\end{minipage}%
\begin{minipage}[b]{0.3\textwidth}
\centering
\includegraphics[scale=0.3]{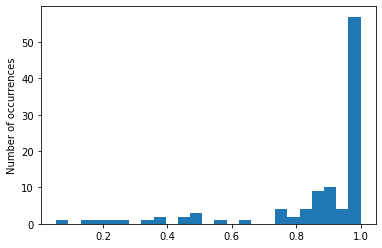}\\
\end{minipage}%
\caption{Histograms of the priorities of \emph{RPG}, \emph{Racing} and \emph{Roguelike Deckbuilder}.}
\label{fig:highPrio}
\end{figure*}

Before defining the taxon of High priority tags, let us first consider priority histograms of tags that clearly should be in that taxon. Three are depicted in Figure~\ref{fig:highPrio}. We observe that these histograms roughly look like Dirac functions centred at $1$. We want to define our High priority tags taxon as the set of tags whose histograms look similarly to those three. To allow for some wiggle room, we decided to define the taxon as the set of tags whose priority median is sufficiently high, at least $0.574803$, and whose maximum value in the histogram is obtained for a priority of at least $0.765644$. Both thresholds were decided using our own expertise of video games. The median threshold was chosen so that the tag \emph{Twin Stick Shooter} would be in that taxon, as its priority median is exactly that value, but not \emph{Underground} and \emph{Conversation} that were following. The maximum value threshold was chosen such that \emph{Tabletop} would be a High priority tag, but not \emph{2.5D} and \emph{Lore-Rich} that were following. With this definition, we have $155$ High priority tags. By looking at these tags, it seems that nearly all of them are about genre. We discuss this in further detail in Subsection~\ref{subsec:taxonomy_low}.

\subsubsection{Medium priority tags}

\begin{figure*}
\centering
\begin{minipage}[b]{0.3\textwidth}
\centering
\includegraphics[scale=0.3]{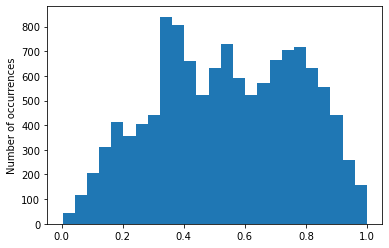}\\
\end{minipage}%
\begin{minipage}[b]{0.3\textwidth}
\centering
\includegraphics[scale=0.3]{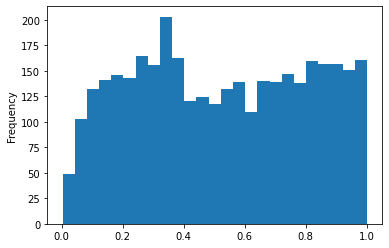}\\
\end{minipage}%
\begin{minipage}[b]{0.3\textwidth}
\centering
\includegraphics[scale=0.3]{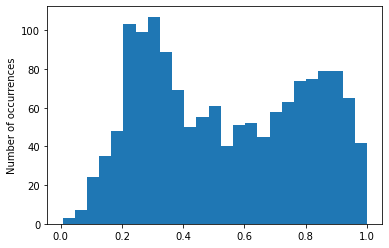}\\
\end{minipage}%
\caption{Histograms of the priorities of \emph{2D}, \emph{Open World} and \emph{PvE}.}
\label{fig:mediumPrio}
\end{figure*}

\begin{figure*}
\centering
\begin{minipage}[b]{0.3\textwidth}
\centering
\includegraphics[scale=0.3]{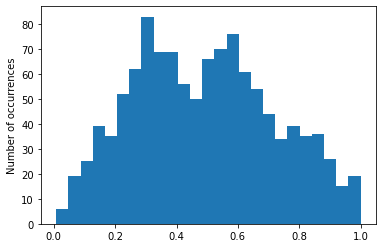}\\
\end{minipage}%
\begin{minipage}[b]{0.3\textwidth}
\centering
\includegraphics[scale=0.3]{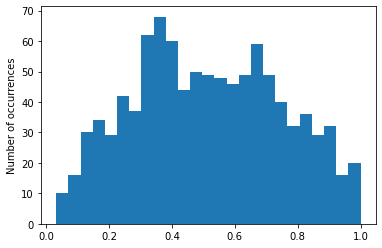}\\
\end{minipage}%
\begin{minipage}[b]{0.3\textwidth}
\centering
\includegraphics[scale=0.3]{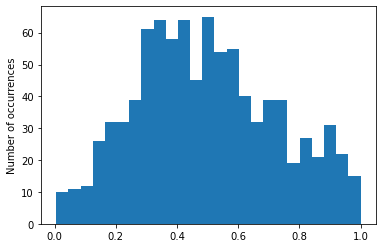}\\
\end{minipage}%
\caption{Histograms of the priorities of \emph{1990's}, \emph{Military} and \emph{Robots}.}
\label{fig:mediumPrio_2D}
\end{figure*}

We have one last taxon of higher rank for all of the remaining $192$ tags. When looking at the histograms, we distinguish three types of shapes. Figure~\ref{fig:mediumPrio} depicts them. We have curves that look like Gaussian curves, as with \emph{2D}, curves that look rather constant, like the one of \emph{Open World}, and curves with two peaks, like the one of \emph{PvE}. However, it seems that these types of shapes do not procure any insightful meaning. Indeed, one can see in Figure~\ref{fig:mediumPrio_2D} three histograms of other Medium priority tags. Observe that they are all similar to the one of \emph{2D}. However, we do not see any similarity between \emph{2D}, \emph{1990's}, \emph{Military} and \emph{Robots} beyond the fact that they are Medium priority tags. It seems to us that in this case, the shape of the histogram does not bring us any further insight.

\subsection{The taxa of lower rank}
\label{subsec:taxonomy_low}

In this subsection, we divide taxa of higher rank into several taxa. Our most interesting comments concerns High priority tags, thus we focus on their study.

\subsubsection{The taxon of High priority tags}
\label{subsubsec:high_priority}

Interestingly, all High priority tags related to gameplay are genre names: \emph{Adventure}, \emph{RPG}, \emph{Farming Sim} and so on. Furthermore, nearly all tags commonly understood as genre tags are in the High priority tags taxon. We consider this finding to be one of the main contributions of this paper. We can now define ``genre tags'' as ``gameplay related tags with high priority''. Also, it provides empirical evidence for Hypothetical Stylised Fact~\ref{fact:genre}: Genre is indeed what players think about first when describing a game.

Some tags are assigned to fewer than $100$ games. We believe that this is too few games for qualifying those tags as ``genres''. We create a taxon for those tags that have high priority median, but that are assigned to too few games. Those are: \emph{BMX}, \emph{Cycling}, \emph{Motocross}, \emph{Skateboarding}, \emph{Tennis}, \emph{Baseball}, \emph{Lemmings}, \emph{Wrestling}, \emph{Mini Golf}, \emph{Social Deduction}, \emph{Warhammer 40K}, \emph{Roguevania}, \emph{Medical Sim}, \emph{LEGO}, \emph{Outbreak Sim}, \emph{Pinball}, \emph{Spelling}, \emph{Golf}, \emph{Jet} and \emph{Action RTS}. 

Steam products are not limited to video games, strictly defined. Other products are also available, and their corresponding tags are all in the High priority tags taxon. This is unsurprising, as if these products were deemed to be games, then the following High priority tags would undoubtedly qualify as game genres: \emph{Utilities}, \emph{Audio Production}, \emph{Video Production}, \emph{Animation \& Modeling}, \emph{Design \& Illustration}, \emph{Photo Editing}, \emph{Software Training} and \emph{Web Publishing}. On a related note, we also have the tags \emph{Game Development} and \emph{Programming} which are more about making than playing video games. Those tags we put on a separate taxon. In Subsection~\ref{subsec:syno}, we provide further evidence to support our decision to treat those tags as an independent taxon.

In total, we divide the taxon of High priority games into four taxa: game genres (which are gameplay related tags), tags assigned to too few games, tags related to products that are not video games, and the miscellaneous tags. In this last taxon we find: \emph{Free to Play}, \emph{Indie}, \emph{Early Access}, \emph{Massively Multiplayer}, \emph{e-sports}, \emph{Sexual Content}, \emph{Nudity}, \emph{LGBTQ+}, \emph{Dinosaurs}, \emph{Mechs}, \emph{Cats}, \emph{Experimental}, \emph{Noir}, \emph{Lovecraftian} and \emph{Western}. 

We list the learnings we obtain from these tags being High priority tags. Unsurprisingly, players care about whether the game is free. We indeed expect players to mind how much they have to pay for a game, which is why \emph{Free to Play} is a High priority tag. Secondly, the tags \emph{Indie}, \emph{Early Access}, \emph{Massively Multiplayer} and \emph{e-sports} provide a lot of information about the type of game, even though those are not genres. Players know better how much money was invested into the game, whether the game is entirely finished, whether it involves a lot of interactions with other players, and whether it is made for e-sports. This tag in itself is full of information. It implies that the game is hard to master, that it has a multiplayer mode, and that it is competitive. The presence of the tags \emph{Sexual Content} and \emph{Nudity} is also not surprising, as they can be important filters for players to know whether they want to play a game. Next, we have the tags \emph{LGBTQ+}, \emph{Dinosaurs}, \emph{Mechs} and \emph{Cats}\footnote{Interestingly, players are more interested in cats than in dogs. Indeed \emph{Cats} has a median priority of $0.60$ with a maximum value in the histogram at $0.96$, whereas for the tag \emph{Dog} those values are $0.56$ and $0.45$, respectively.} which are themes that players think to be so important that they are worth mentioning right away when describing a game. Similarly, \emph{Experimental}, \emph{Noir}, \emph{Lovecraftian} and \emph{Western} are ambiances of high importance for players when describing a game. We believe that game developers can benefit from knowing that these themes and ambiances matter particularly to players. 

\subsubsection{The taxon of Medium priority tags}
\label{subsubsec:false_medium}
As we said, our most interesting comments are for High priority tags. In particular, we showed in Figure~\ref{fig:mediumPrio_2D} that the priority histogram was not sufficient for classifying the Medium priority tags into meaningful taxa. Thus, for the rest of the section, we only discuss why a few tags widely regarded as genres were put into the Medium priority tags taxon. Those tags are: \emph{FPS}, \emph{Shooter}, \emph{Fighting}, \emph{Stealth}, \emph{Hack and Slash}, \emph{Survival}, \emph{Survival Horror}, \emph{Horror}, \emph{MOBA}, \emph{4X}, \emph{RTS}, \emph{Grand Strategy}, \emph{Trading Card Game}, \emph{Match 3}, \emph{Hidden Object} and \emph{MMORPG}. We still believe that those tags should be considered as genre, and explain why they were misclassified. Concerning \emph{FPS}, \emph{Shooter}, \emph{Fighting}, \emph{Stealth}, \emph{Hack and Slash}, \emph{Survival}, \emph{Survival Horror} and \emph{Horror}, they have relatively low priority medians because they are often used by players to say that some elements from these genres appear in a game. For instance, if it is possible to fight in a game, then the tag \emph{Fighting} will surely appear with medium priority. \emph{Fighting} is not the genre of the game, but it corresponds here to a gameplay element. Let us take another example with \emph{FPS} and \emph{Shooter}. The most assigned tags to the game \emph{Fallout 4} are \emph{Open World}, \emph{Post-apocalyptic}, \emph{Exploration} and \emph{RPG}. In our opinion, those tags indeed describe well this game. With a lower priority, we find also the tags \emph{Shooter} and \emph{FPS}. It is true that there are FPS elements in the game, which occur for instance when the player is aiming, therefore it is understandable that the tag appears with a medium priority. The takeaway is that tags may have different purposes and connotations. There are tags that might be used as genres, but also sometimes used to merely describe some gameplay elements. If it were a common practice to distinguish the two things and have a tag \emph{FPS - genre} and a tag \emph{FPS - gameplay}, we believe that the former would be put into the High priority tags taxon while the latter would be rightly classified as a Medium priority tag. It was most surprising to us to discover that \emph{MOBA} is in that taxon. We expected it to only appear as a High priority tag. However, its priority median is quite low, with a value of $0.48$, making it even closer to Low priority tags than High priority tags. After investigation, it appears that the tag is often assigned with a medium priority to team-based multiplayer games, even when they do not belong to the MOBA genre. It seems that players and the industry have a different understanding of the tag.

Concerning the remaining tags, \emph{4X}, \emph{RTS}, \emph{Grand Strategy}, \emph{Trading Card Game}, \emph{Match 3}, \emph{Hidden Object} and \emph{MMORPG}, they seem to be overshadowed by other tags preferred by players. For instance \emph{4X}, \emph{RTS} and \emph{Grand Strategy} are overshadowed by \emph{Strategy}, \emph{Turn-Based} and \emph{Real-Time}. \emph{Trading Card Game} often appears after \emph{Card Game} and \emph{Card Battler}. \emph{Match 3} and \emph{Hidden Object} are overshadowed by \emph{Puzzle}. Likewise, \emph{MMORPG} is regularly chosen only after \emph{RPG}, \emph{Multiplayer} and \emph{Massively Multiplayer}. This is why, although they have a high priority median, their occurrence peak in the histogram is not close to a priority value of $1$.

\subsection{List of genres comparison}

Using our data-driven approach, we have now obtained a list of genre tags. It can be found in Appendix~\ref{app:list_genres}. Those are mainly in the High priority taxon, and we presented in Paragraph~\ref{subsubsec:false_medium} a list of tags that should still be considered as genre tags, although they were classed as Medium priority tags. We compare our list to Li's~\cite{li2020towards} and to the one in Steamworks~\cite{steamworks2021steam} as they are both made from Steam tags. It would be interesting to compare ours to all genre lists. However, for the sake of brevity, we concentrate solely on those two lists.

Li's list, although also being made from a data-driven approach, serves a different role than ours~\cite{li2020towards}. His aim was to extract from Steam tags a reasonably small list of genre tags that would allow one to characterise all games, whereas our goal is to find an extensive list of genres. Consequently, Li's list contains $29$ genres, and ours contains $127$ tags. Still, the comparison is interesting as there are a few tags which are in Li's list but not in ours. Those are: \emph{Soccer}, \emph{Resource Management}, \emph{Music} and \emph{Classic}. For each genre, Li associates a group of correlated gameplay tags. Concerning \emph{Soccer}, the group contains \emph{Soccer}, \emph{Football} and \emph{Sports}. This choice seems strange to us, for then according to Li's list all sports game would be considered as soccer games. In our genre list, we have the tag \emph{Sports} instead of \emph{Soccer}, which seems more reasonable. The same happens with \emph{Resource Management} that we would replace by \emph{Management} and \emph{Music} that we would replace by \emph{Rhythm}. Concerning \emph{Classic} in Li's list, its presence follows from the fact that Li considers \emph{Classic} and \emph{Cult Classic} to be related to gameplay. We disagree with that statement, although we cannot substantiate our belief with data.

Similar to ours, Steamworks' list of genre tags strives for comprehensiveness~\cite{steamworks2021steam}. Our list contains $127$ tags whereas theirs contains $140$. The two lists share $112$ tags in common, which indicates that they mostly agree with each other. Let us start with the $15$ tags that are in our list but not in Steamworks'. It is worth noting that despite Steamworks appearing to have all of the Steam tags (whether genre-related or not), some tags are actually absent. The genre tags in our list that do not even appear on Steamworks are: \emph{Cooking}, \emph{Creature Collector}, \emph{Party Game}, \emph{Puzzle-Platformer} and \emph{Roguelike Deckbuilder}. More interestingly, there are a few tags that are High priority tags related to gameplay (which we thus consider as genre tags), which are not considered as genre tags by Steamworks. We provide the list here: \emph{Archery}, \emph{Automation}, \emph{Boxing}, \emph{Deckbuilding}, \emph{Fishing}, \emph{Hunting}, \emph{Naval Combat}, \emph{Vehicular Combat}, \emph{Otome} and \emph{Parkour}. All those tags were classed as Features in Steamworks instead of genres, except for \emph{Otome} and \emph{Parkour} which are classed as Themes \& Moods. We think that the fact that those are High priority tags supports the idea of considering them as genre tags.

We present the $28$ tags that are in Steamworks' list of genre tags but not in ours. Some are High priority tags about gameplay that we removed in Paragraph~\ref{subsubsec:high_priority} because they were assigned to fewer than $100$ games. Indeed, it seems to us that tags assigned to too few games cannot really define a genre. Those tags are: \emph{BMX}, \emph{Baseball}, \emph{Cycling}, \emph{Golf}, \emph{Medical Sim}, \emph{Mini Golf}, \emph{Motocross}, \emph{Outbreak Sim}, \emph{Roguevania}, \emph{Pinball}, \emph{Skateboarding}, \emph{Spelling}, \emph{Tennis} and \emph{Wrestling}. There are also a few High priority tags that we did not put into our list of genre tags because we deem that they are not clearly enough related to gameplay. Those are \emph{Programming}, \emph{e-sports} and \emph{Experimental}. The remaining tags are all Medium priority tags, and are therefore not part of our genre lists. Our findings indicate that those are not words that first come to players mind when describing a game, and supports the idea of not considering them as genres. The tags are: \emph{Basketball}, \emph{Bowling}, \emph{Football}, \emph{Hacking}, \emph{Hockey}, \emph{Open World}, \emph{Skating}, \emph{Skiing}, \emph{Snowboarding}, \emph{Soccer} and \emph{Investigation}.

\section{Two meronomies of Steam tags}
\label{sec:meronomy}

In this section, we propose two meronomies of some Steam tags. In Section~\ref{sec:taxonomy}, we classified the tags into taxa, where two tags in the same taxa share some properties (related to their priority distributions). In Section~\ref{sec:meronomy}, we use totally different methodologies, for different goals. In the resulting meronomies, we say that a tag $A$ is a part of a broader tag $B$, or equivalently that $A$ is a meronym of $B$.

The first meronomy we present focuses on the tags with the broadest meanings. We aim at finding the most general tags, and to see how they relate to each other. In particular, we focus on the tags at the top of the meronomy, the ones that are parts of no others, but all others tags are parts of them. We name them the ``capital tags''. We claim that the capital tags are interesting because they form the smallest set of words that one needs to describe all video games (without going into detail). 

The second meronomy aims at finding synonymous tags. We believe that one can reduce the number of Steam tags without loosing too much information, as many tags share similar meanings. Our aim is to find those groups of tags, and for each one to find one representative.

\subsection{The method}
\label{subsec:method}

The method starts the same for both meronomies. The idea is to look at pairs of tags $A$ and $B$ that are correlated enough, and then check whether $A$ or $B$ appears more often. If $B$ appears more often, we declare that $A$ is a meronym of $B$. Intuitively, $A$ should indeed be a part of $B$, as they are correlated and $B$ appears more often, ergo has a broader meaning.

For each game $G$, we have an entry with $427$ values (the number of Steam tags), where the value is the priority of the corresponding tag for the game $G$. As there are at most $20$ tags with positive priority for any given game, it implies that at least $407$ values are equal to $0$. We compute the Pearson correlation coefficient for all pairs of tags. A natural idea would be to take all pairs with correlation above some threshold and be done with it. The issue is that many tags are assigned to very few games. Thus, their priority is $0$ for virtually all games. This induces a bias, and two of such tags will have a strong correlation even though they are hardly related. To tackle this issue, we consider all pairs of tags $A$ and $B$ with correlation above $0.1$, and then compute what we call their \emph{local Pearson correlation coefficient}. It is equal to the Pearson correlation coefficient, but it is computed only on games to which at least one of $A$ or $B$ is assigned. 

For the first meronomy, we keep only the pairs with local Pearson correlation coefficient above $-0.7$, over all pairs with (global) Pearson correlation above $0.1$. For the second meronomy, we take all pairs with local Pearson coefficient above $-0.6$, while keeping a global Pearson correlation threshold of $0.1$. Let us motivate the choice of these thresholds. For the global threshold, the lower it is, the better for the soundness of the results. However, getting it lower increases a lot the number of pairs of tags for which we have to compute the local Pearson correlation. As we have $428$ tags, there are $\binom{428}{2}=91378$ pairs to consider. This is computationally quite expensive, which is why we only consider pairs with correlation above $0.1$. There are $2197$ of those pairs. As mentioned earlier, we have concerns with the global correlation approach in that it may produce some false positives (pairs with high correlation that are not closely related). However, we do not anticipate the occurrence of a false negative, that is a pair of tags with similar meanings but low correlation.

For the local threshold of the first meronomy, the value $-0.7$ may seem extremely low. However, as we do the computation only on the games to which at least one of the tags is assigned, this diminishes drastically the correlation value. Indeed, there are proportionally much more games to which only one tag is assigned, since we removed all games to which none are assigned. Thus, with a local threshold of $-0.7$, we are actually considering only $652$ pairs of tags out of the $2196$ that have global correlation above $0.1$. Getting the threshold even lower would make us consider pairs that are actually not so much related, according to our own video game expertise. We present the result in Subsection~\ref{subsec:capital}.

We study what happens when we set the local threshold to $-0.6$ in Subsection~\ref{subsec:syno}. There are only $361$ pairs with local correlation above this threshold. The choice of the value was again made using our own video game expertise. We chose the lowest value such that the pairs were indeed synonymous in our opinion.

\subsection{The capital tags}
\label{subsec:capital}

\begin{figure}
    \centering
    \includegraphics[scale=0.8]{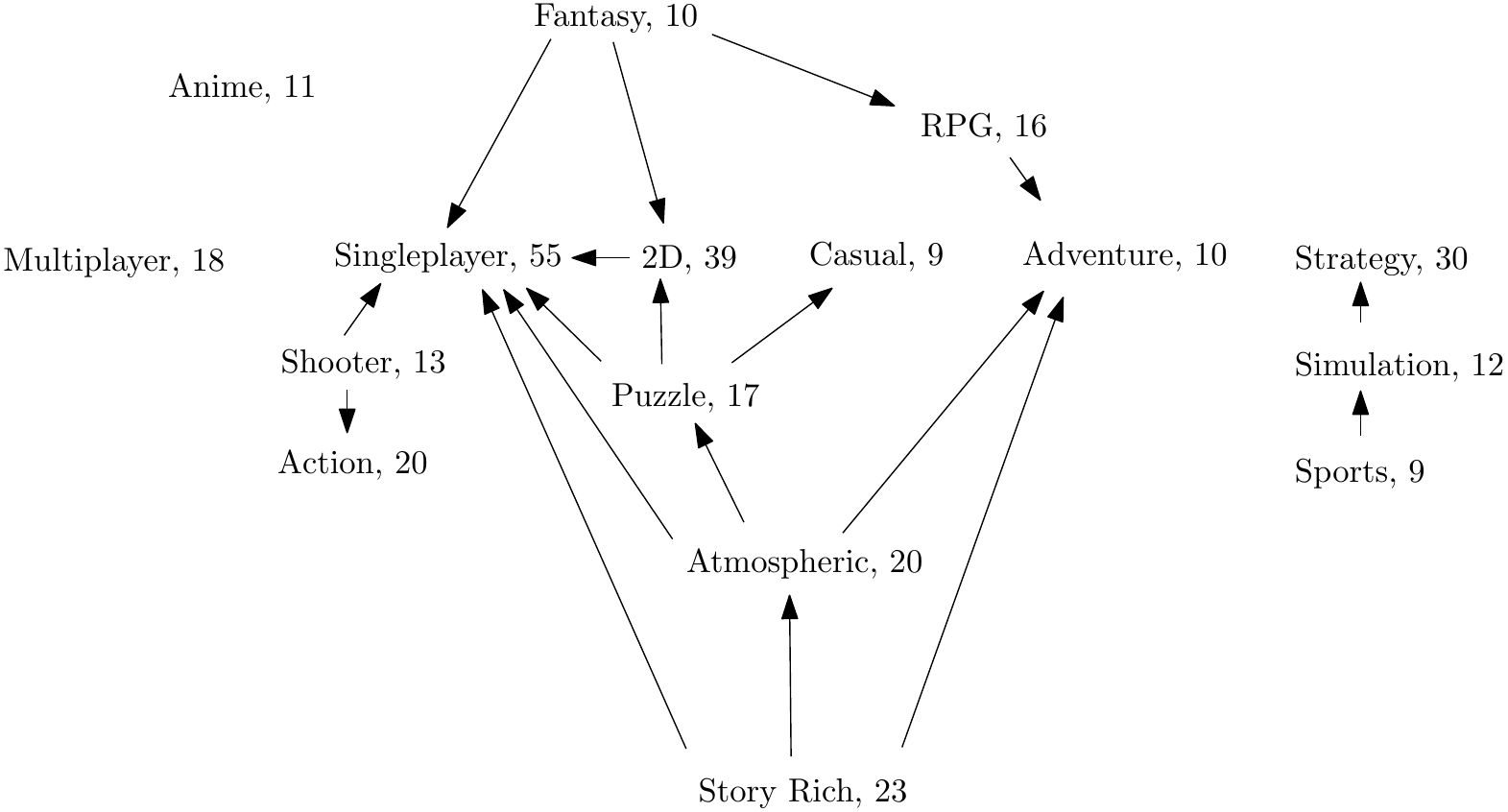}
    \caption{The oriented subgraph induced by the tags with at least 9 incoming edges. For each tag, we write its name and the number of incoming edges in the original graph.}
    \label{fig:oriented}
\end{figure}

Now that we have a list of $652$ pairs of tags with high correlation, we compute for each pair $A$, $B$ whether $A$ or $B$ appears more. We construct an oriented graph with the tags as vertices and an edge from tag $A$ to tag $B$ if $B$ occurs more often than $A$. This graph is unreadable as there are too many vertices and edges. However, we are mainly interested in the vertices that have the most incoming edges, as they correspond to tags with the broadest meanings. Figure~\ref{fig:oriented} shows the subgraph induced by the vertices $V$ with at least $9$ incoming edges (that is the subgraph where the vertex set is $V$, and we show only edges between vertices in $V$). We observe that there are seven vertices with no outgoing edges, that we name the \emph{capital tags}. One might observe that the graph shown in Figure~\ref{fig:oriented} does not perfectly correspond to a meronomy. We tackle this issue at the end of the subsection, as we want first to focus on the capital tags.

\begin{definition}\normalfont
The seven capital tags are \emph{Multiplayer}, \emph{Singleplayer}, \emph{Action}, \emph{Casual}, \emph{Adventure}, \emph{Strategy} and \emph{Anime}. They correspond to tags that are correlated with a lot of other tags, and that occur more than them.
\end{definition}

The name is a reference to the seven capital sins from Christian teachings. Thomas Aquinas uses the word ``capital'' for the following reason: Accordingly a capital vice is so called, in the first place, from ``head'' taken in the proper sense, and thus the name ``capital'' is given to a sin for which capital punishment is inflicted. It is not in this sense that we are now speaking of capital sins, but in another sense, in which the term ``capital'' is derived from head, taken metaphorically for a principle or director of others. In this way a capital vice is one from which other vices arise, chiefly by being their final cause~\cite{aquinas1920summa}.

We posit that a similar phenomenon exists with tags, where the capital tags serve as the directors of the other Steam tags, much like how the capital vices act as the head of other vices. The seven capital tags encompass all others, as stated in the following hypothetical stylised fact:

\begin{stylised}\label{fact:capital_parts}
All tags are parts of the capital tags.
\end{stylised}

By construction, our seven capital tags are tags that are correlated with a lot of other tags, and that occur more often than them. This is already some justification for Hypothetical Stylised Fact~\ref{fact:capital_parts}. We add some more evidence at the end of the subsection when we define a meronomy of the tags in Figure~\ref{fig:oriented}. Before that, we state and establish another hypothetical stylised fact. First, we need to define the following concept:

\begin{definition}\normalfont
We say that a game $G$ is covered by a set of tags~$\mathcal{T}$ if at least one of the tags in $\mathcal{T}$ is assigned to~$G$.
\end{definition}

\begin{stylised}\label{fact:capital}
The capital tags form a very small group of tags such that nearly all games are covered by this set.
\end{stylised}

Hypothetical Stylised Fact~\ref{fact:capital}, although being quantitative in essence, is not perfectly mathematically defined. One reason for this is that the Steam tag are assigned by players. So all rules and observations should suffer from a few counterexamples, as the data is significantly noisy. On another note, it is computationally too hard to check Hypothetical Stylised Fact~\ref{fact:capital}. If one wanted to check all sets of seven tags for instance, that would be $\binom{428}{7}\approx4.97 \cdot 10^{14}$ sets of tags to test.

\begin{table}
\resizebox{\columnwidth}{!}{%
\begin{tabular}{l|ll|ll|}
\cline{2-5} & \multicolumn{2}{l|}{percentage of covered games}   & \multicolumn{2}{l|}{percentage of covered games by this tag and the ones above} \\ \hline
\multicolumn{1}{|l|}{tag}          & \multicolumn{1}{l|}{over all games} & over games with exactly $20$ tags & \multicolumn{1}{l|}{over all games}  & over games with exactly $20$ tags  \\ \hline
\multicolumn{1}{|l|}{\emph{Singleplayer}} & \multicolumn{1}{l|}{46}             & 67                                  & \multicolumn{1}{l|}{46}                & 67      \\ \hline
\multicolumn{1}{|l|}{\emph{Action}}       & \multicolumn{1}{l|}{45}             & 54                                  & \multicolumn{1}{l|}{70}                & 85      \\ \hline
\multicolumn{1}{|l|}{\emph{Casual}}       & \multicolumn{1}{l|}{43}             & 42                                  & \multicolumn{1}{l|}{84}                & 92        \\ \hline
\multicolumn{1}{|l|}{\emph{Adventure}}    & \multicolumn{1}{l|}{42}             & 51                                  & \multicolumn{1}{l|}{90}                & 95         \\ \hline
\multicolumn{1}{|l|}{\emph{Strategy}}     & \multicolumn{1}{l|}{21}             & 26                                  & \multicolumn{1}{l|}{93}                & 98           \\ \hline
\multicolumn{1}{|l|}{\emph{Multiplayer}}  & \multicolumn{1}{l|}{13}             & 22                                  & \multicolumn{1}{l|}{93}                & 98      \\ \hline
\multicolumn{1}{|l|}{\emph{Anime}}        & \multicolumn{1}{l|}{9}              & 14                                  & \multicolumn{1}{l|}{94}                & 98          \\ \hline
\end{tabular}
}
\caption{How the capital tags cover nearly all games}
\label{tab:capital}
\end{table}

One can see in Table~\ref{tab:capital} that our choice of capital tags does cover nearly all games. We show the percentage of games covered by each of the capital tag. We also show the percentage for the set consisting of the tag of the line and all the ones above. The tags are sorted by decreasing percentage of games covered. We also do the same when considering only games with $20$ tags. Games with fewer tags can be misleading, as it may simply be that not enough players assigned tags, and if more players did, then surely some of them would have assigned a capital tag. We observe that Table~\ref{tab:capital} strongly supports Hypothetical Stylised Fact~\ref{fact:capital}.

As argued, it is not computationally possible to test all sets of seven games. Still, we present another set that could have been a fitting candidate: \emph{2D}, \emph{3D}, \emph{2.5D}, \emph{Isometric}, \emph{Third Person}, \emph{First-Person} and \emph{Top-Down}. All games are either in 2D, in 3D, or in some mix of the two: 2.5D or isometric. We even add \emph{Third Person} and \emph{First-Person} which imply that the game is in 3D, but give more information about the representation. The same holds with \emph{Top-Down} and \emph{2D}. Therefore, one might expect to cover all games with these seven tags. This is not the case, as only $47$\% of the games are covered, and only $85$\% of the games with $20$ tags are covered. This again supports Hypothetical Stylised Fact~\ref{fact:capital}, as it substantiates the claim that the capital tags are very few while covering essentially all games.

For the sake of curiosity, we extended the capital tags with as few tags as possible in order to cover even more games. This was done by hand by looking at which games were not covered yet, and finding what tags were assigned to most of them. By adding the tags \emph{Arcade}, \emph{RPG}, \emph{Simulation}, \emph{Survival}, \emph{Puzzle} and \emph{Horror} to the capital tags, we cover $96.73$\% percent of games, and $99.78$\% percent of games with $20$ tags. Unsurprisingly, several of those new tags can be found in Figure~\ref{fig:oriented}: \emph{Puzzle}, \emph{RPG} and \emph{Simulation}.

We make a few side-remarks that we deem interesting. First, we observe that for six of the capital tags, the percentage of games covered is less than the one of games with $20$ tags covered. The only one for which it decreases is \emph{Casual}, which goes from $43$\% to $42$\%. Moreover, the increase for the six others is quite different from tag to tag. The biggest increase in terms of scale factor happens with \emph{Multiplayer} (scale factor of $1.69$). Intuitively, the games with $20$ tags are the most mainstream. This supports the idea that mainstream games on Steam more often offer multiplayer modes, and that they tend to be less casual. We also note that the second biggest increase is with \emph{Singleplayer} (scale factor of $1.46$). This tells us that even though \emph{Multiplayer} and \emph{Singleplayer} are classified in the Low Priority taxon (see Subsection~\ref{subsec:taxonomy_high}), they are actually assigned to many games, and would perhaps be assigned to all games if all games had $20$ tags assigned to them. Therefore, low priority tags should not be thought of as of less importance. They are not what players think about first when describing a game, but may still be a feature that should be mentioned at a later step.

Recall that a meronomy deals with part-whole relationships. Mathematically, a meronomy can be seen as a partially ordered set (poset). A partial order, denoted by $\leq$, must satisfy three properties:

\begin{enumerate}
    \item Transitivity: if $A\leq B$ and $B\leq C$ then $A\leq C$,
    \item Reflexivity: we have $A \leq A$,
    \item Antisymmetry: if $A\leq B$ and $B\leq A$, then $A=B$.
\end{enumerate}

We can see in Figure~\ref{fig:oriented} that the relation depicted is not perfectly a partial order. Indeed, let us say that $A$ is a part of $B$, denoted by $A\leq B$, if there is an arrow from $A$ to $B$. The graph in Figure~\ref{fig:oriented} looks a lot like a Hasse diagram. A Hasse diagram of a poset $\leq$ is a representation where there is an edge from $A$ to $B$ if $A\leq B$ and there is no $C$ such that $A\leq C \leq B$. This is a useful and concise way of representing posets. For the graph in Figure~\ref{fig:oriented} to be a Hasse diagram, one simply has to remove the edges from \emph{Puzzle}, \emph{Atmospheric}, \emph{Story Rich} and \emph{Fantasy} to \emph{Singleplayer}. Thus, we are implying that we consider \emph{Sports} to be a part of \emph{Simulation}, which itself is a part of \emph{Strategy}. Likewise, we say that \emph{Fantasy} is a part of \emph{RPG}, which itself is a part of \emph{Adventure}.

Let us mention some other relationships, not depicted in Figure~\ref{fig:oriented}, but that are present in the larger graph (which we did not represent as there are too many edges to be readable). We have arrows from \emph{Co-op} to \emph{Multiplayer} and \emph{Action}. There is an arrow from \emph{Cute} to \emph{Colorful}, itself being connected to \emph{2D} and \emph{Casual}. \emph{Management} is connected to \emph{Singleplayer} and \emph{Simulation}. Finally, we want to mention \emph{PvP} being connected to \emph{Multiplayer}. As all these relations are quite sensible, this supports our idea of looking at that graph as the Hasse diagram of some poset, or equivalently to establish a meronomy of those tags.

We now have evidence for Hypothetical Stylised Fact~\ref{fact:capital_parts}: All tags are parts of the capital tags. By only adding or removing a few edges, we have obtained a meronomy for the tags of Figure~\ref{fig:oriented} that could be extended further to the other tags. In this meronomy, we have seen that the capital tags are the ones at the top: All other tags are meronyms of the capital tags.

Finally, we make a concluding remark about the capital tags. Four of them are game genres: \emph{Action}, \emph{Adventure}, \emph{Strategy} and \emph{Casual} (although it is debatable whether it is a genre, this is how it is considered in Steamworks' and our genre list~\cite{steamworks2021steam}). Two other tags give important information about the game, even though they are not genres: \emph{Singleplayer} and \emph{Multiplayer}. We suspect that the remaining tag, \emph{Anime}, is actually an error. First, it is the tag among the capital tags that covers the smallest number of games. Secondly, the $11$ tags that have an edge directed to it in the graph are: \emph{JRPG}, \emph{Visual Novel}, \emph{Otome}, \emph{Dating Sim}, \emph{Romance}, \emph{Female Protagonist}, \emph{Sexual Content}, \emph{NSFW}, \emph{Hentai}, \emph{Nudity} and \emph{Mature}. We observe that the five latter tags refer to sexual content, and share a similar meaning. This might induce a bias, giving to much weight to tags related to sexual content, therefore wrongly putting \emph{Anime} as a tag with many incoming edges in the graph.

\subsection{The synonymous tags}
\label{subsec:syno}

As detailed in Subsection~\ref{subsec:method}, we now use a local correlation threshold of $-0.6$. We deem the pairs of tags that have a local correlation that high to be synonymous. In Subsection~\ref{tab:capital}, all edges were oriented from a tag to another: there was an arrow from $A$ to $B$ if $B$ appears much more often than $A$. In this subsection, we are also interested in knowing when tags appear mostly together. Indeed, this would be the best examples of tags being synonymous. When tags appear mostly together, we draw a  \emph{mutual} edge: an arrow that points towards both ends. To decide what kind of edge to draw, we do the following: For the tags $A$ and $B$, we compute $X_A$, $X_B$, $X_{A\cup B}$ and $X_{A\cap B}$ which denote how many times $A$ was assigned to a game but not $B$, how many times $B$ was assigned to a game but not $A$, how many times at least one of $A$ or $B$ was assigned, and how many times both were assigned, respectively. We define $r_A:=X_A/X_{A\cup B}$, $r_B:=X_B/X_{A\cup B}$ and $r_{A\cap B}:=X_{A\cap B}/X_{A\cup B}$. If the maximum of these three ratios is $r_A$, then we draw an arrow from $B$ to $A$. If it is $r_B$, then we draw an arrow from $A$ to $B$. If it is $r_{A\cap B}$, then we draw a mutual arrow between $A$ and $B$.

Before running the program on the tags, we remove a few of them. The reason for this is that there are tags which are correlated with a lot of other tags, and that are assigned to many games. Those are the ones we mentioned in Subsection~\ref{subsec:capital}. We remove the capital tags in addition to \emph{2D}, \emph{Shooter}, \emph{Puzzle}, \emph{Atmospheric}, \emph{Simulation}, \emph{Story Rich} and \emph{Fantasy} (which are amongst the other tags with the highest number of edges oriented towards them in Figure~\ref{fig:oriented}). Keeping those tags would distort the result and show only correlation between those tags and few others.

Also, we use the information we obtained in Subsection~\ref{subsec:taxonomy_high}. Recall that we classified the tags into three taxa of higher rank: High priority tags, Medium priority tags and Low priority tags. In the oriented graph of this subsection, we draw dotted red arrows for edges where the two involved tags are not from the same taxon. Indeed, not being in the same taxon may be a hint that these tags are not synonymous after all, since they are not used in the same way by players, although being highly correlated.

\begin{figure}
    \centering
    \includegraphics[scale=0.76]{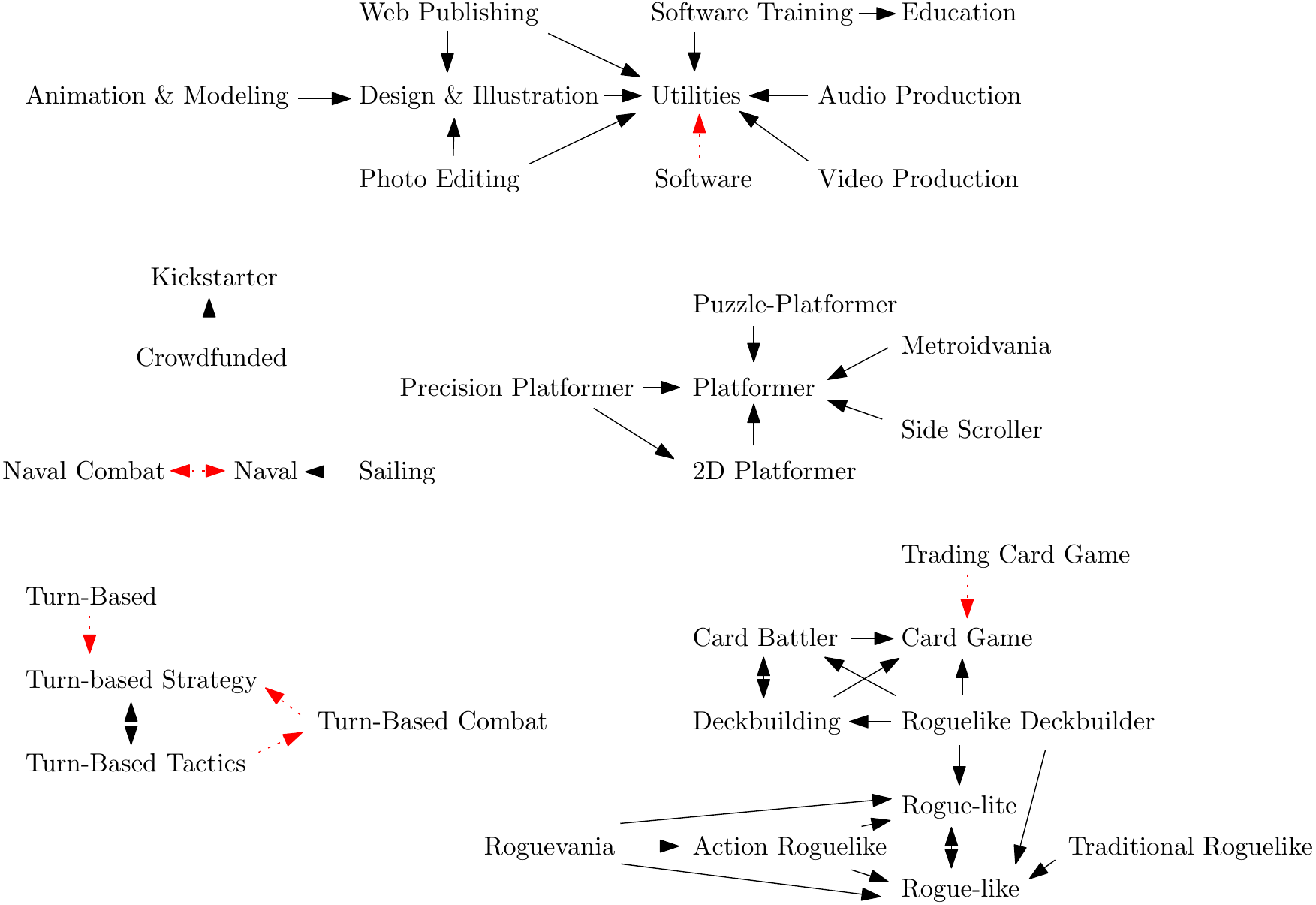}
    \caption{Some synonymous tags. Arrows are dotted red if the tags belong to different taxa of higher rank.}
    \label{fig:synonyms}
\end{figure}

Some synonymous tags are shown in Figure~\ref{fig:synonyms}. Due to lack of space, we only depict the examples we find most meaningful. As a first simple example, we have an edge from \emph{Crowdfunded} to \emph{Kickstarter}. As wanted, the meanings of those two tags are very close. Let us consider the subgraph induced by \emph{Naval Combat}, \emph{Naval} and \emph{Sailing}. The arrow from \emph{Sailing} to \emph{Naval} is natural, as the meanings are close and \emph{Sailing} implies \emph{Naval}. However, \emph{Naval Combat} and \emph{Naval} are not so much synonymous. Indeed, one is understood as a game genre, whereas the other merely indicates that a game involves ships. This is why the information from Subsection~\ref{subsec:taxonomy_high}, depicted by the dotted red arrow, does matter. Although it is inevitable that \emph{Naval Combat} and \emph{Naval} be correlated, it seems wrong to treat them as synonyms. Likewise, the arrow from \emph{Turn-Based} to \emph{Turn-Based Strategy} is treacherous, as one is a gameplay element, and the other is a game genre. More generally, we believe that no red arrow should be taken into account. We make one exception for the edge from \emph{Trading Card Game} to \emph{Card Game}. Indeed, we argued in Subsection~\ref{subsec:taxonomy_low} that \emph{Trading Card Game} was one of the few tags that are misclassified by our method, and that it should actually be treated as a genre. On another note, we have a group of tags related to non-games, that we mentioned in Subsection~\ref{subsec:taxonomy_low}. All those tags can be considered as one group of tags related to products offered by Steam that are not games.

When looking at Figure~\ref{fig:synonyms}, a natural idea is to give a name for all these groups of synonymous tags. This name would be a representative of the group, the one that is most general. For instance, the group (\emph{Crowfunded}, \emph{Kickstarter}) would be represented by \emph{Kickstarter}. Similarly, the group of platformer tags would be represented by \emph{Platformer}. But how would one deal with the group that contains card game tags and rogue-like tags? Intuitively, this group should be split in two. Likewise, what about the group of tags not related to games? It would seem natural to have most of those tags represented by \emph{Utilities}, but it is no clear what should happen to \emph{Education}, as the only edge incident to it is incoming, and not outgoing. We detail in the next paragraph a formal explanation on how to deal with these situations, which results in a grouping that is in accordance with our intuition.

Let $G$ be a directed graph with labeled vertices. First merge all mutual edges of $G$. For a mutual edge $e=\{u,v\}$, keep arbitrarily one of the two labels $u$ or $v$. Let $H$ denote the same graph as $G$ with the orientations of the edges removed, resulting in an unoriented graph. Let us denote by $W_1,\dots,W_k$ the vertex sets of the $k$ connected components of $H$. We denote by $V_1,\dots,V_k$ the same vertex sets considered in $G$. Let us consider the subgraph $G_i$ of $G$ induced by $V_i$, for some $1\leq i\leq k$. We say that $G_i$ is \emph{well-oriented} if there exists a unique vertex $v \in V_i$ such that for each $u\neq v$ in $V_i$, there exists a directed path from $u$ to $v$ in $G_i$. Making an abuse of notation, we say that $G$ is \emph{well-oriented} if each $G_i$ is well-oriented.

Going back to the example in Figure~\ref{fig:synonyms}, we see that most connected components are well-oriented. Moreover, for the group of platformer tags, \emph{Platformer} is the unique tag that can be reached via directed paths from all other tags in the group. However, the group with card game tags and rogue-like tags is not well-oriented. Now, what we suggest for splitting into well-oriented groups is the following: Given the graph $G$, remove as few edges as possible to make it well-oriented. To the best of our knowledge, this problem has not been previously studied. However, we believe it is interesting from both a theoretical perspective and for practical applications: an efficient algorithm would be needed for naming the synonym groups. We do not know of any fast algorithm for solving the problem, and wonder whether it is NP-hard. Nonetheless, on the small example we are considering now, we solve the problem by hand. In Figure~\ref{fig:synonyms}, it suffices to remove the edges from \emph{Roguelike Deckbuilder} to \emph{Rogue-lite} and \emph{Rogue-like} to obtain a well-oriented graph. We therefore obtain the group (\emph{Card Battler}, \emph{Deckbuilding}, \emph{Trading Card Game}, \emph{Roguelike Deckbuilder}, \emph{Card Game}) represented by \emph{Card Game}, and the group (\emph{Roguevania}, \emph{Action Roguelike}, \emph{Traditional Roguelike}, \emph{Rogue-lite}, \emph{Rogue-like}) represented by either \emph{Rogue-lite} or \emph{Rogue-like}. Similarly, the group of tags not related to video games is not well-oriented. To make it well oriented, one has to remove the edge from \emph{Software Training} to either \emph{Education} or \emph{Utilities}. Although in this case we solved the problem by hand, it would be very useful to have efficient algorithms for solving this problem for general directed graphs. This would allow us to deal with the synonymous tags on a larger scale.

Finally, we observe that the graph is very similar to the graph of a poset, as was the case in Figure~\ref{fig:oriented}. One could remove the dotted red edges, merge the vertices connected by a mutual edge, and remove the edges we mentioned to make the graph well-oriented. Now it suffices to remove a few edges that give only redundant information, like the one from \emph{Photo Editing} to \emph{Utilities} or the one from \emph{Roguevania} to \emph{Rogue-lite}/\emph{Rogue-like}. Our approach yields a meronomy of synonymous tags, with the topmost tags serving as representatives of their respective groups.

\section{Conclusion}

We proposed several hypothetical facts, such as the notion that players primarily associate games with their genre. To support this claim, we introduced the concept of priority of Steam tags and demonstrated its relationship to players' perceptions of games. Moreover, our approach enabled us to create a data-driven taxonomy that provides a comprehensive list of genres. We found a set of seven tags, the capital tags, which roughly summarise all information contained in the Steam tags. Those are \emph{Multiplayer}, \emph{Singleplayer}, \emph{Action}, \emph{Casual}, \emph{Adventure}, \emph{Strategy} and \emph{Anime}, although we argued why the presence of this last tag in the list is dubious at best. We found the tags \emph{Multiplayer} and \emph{Singleplayer} extremely interesting as they are capital tags, assigned to thousands of games, but are Low priority tags. We showed how some tags can be merged without loosing too much information. We proposed a criterion for finding a representative of a merged group, and ask whether there exists an efficient algorithm for applying this method.

We list some further improvements that can be made to our study. We stated that the notion of priority might need to be refined in order to be more consistent, by taking the example of the tag \emph{Free to Play}. Although considering priority histograms allowed us to define an interesting taxonomy, this method has some limit. In particular for Medium priority tags, we could not find a meaningful way of subdividing this taxon into taxa (see Figure~\ref{fig:mediumPrio_2D}). Maybe some other ideas could allow one to establish a meaningful data-driven classification of Medium Priority tags. Secondly, our method is biased towards games available on Steam, and we acknowledge that this may limit the generalisability of our findings. Also, we note that our computations assume an equal impact for all games. However, according to Orland~\cite{orland2014introducing}, in 2014, over a quarter of registered games on Steam had never been played. Thus, we believe it would be beneficial to consider player engagement metrics, such as the number of players or total hours played, and to weight games accordingly in our computations. We noticed that some Low priority tags are assigned to thousands of games, like \emph{Multiplayer} and \emph{Singleplayer}. We believe that they should be investigated further, to better understand how players think about them. Using the knowledge of capital tags, we developed a method for identifying synonymous tags. It would be intriguing to apply this method in a database where synonymous tags are intelligently merged. Our expectation is that, in such a scenario, \emph{Anime} would no longer be considered a capital tag.


%\begin{acks}
%We thank every Steam player who took time to assign tags to games.
%\end{acks}

\bibliography{bib}

\begin{thebibliography}{10}

\bibitem{adams2006fundamentals}
Ernest Adams and Andrew Rollings.
\newblock {\em Fundamentals of game design (game design and development
  series)}.
\newblock Prentice-Hall, Inc., 2006.

\bibitem{aquinas1920summa}
Thomas Aquinas.
\newblock {\em Summa Theologica: Translated by Fathers of the English Dominican
  Province}.
\newblock Burns Oates \& Washbourne, 1920.
\newblock URL: \url{https://www.newadvent.org/summa/2084.htm#article4}.

\bibitem{arsenault2009video}
Dominic Arsenault.
\newblock Video game genre, evolution and innovation.
\newblock {\em Eludamos: Journal for computer game culture}, 3(2):149--176,
  2009.

\bibitem{baumann2018hardcore}
Florian Baumann, Dominik Emmert, Hermann Baumgartl, and Ricardo Buettner.
\newblock Hardcore gamer profiling: Results from an unsupervised learning
  approach to playing behavior on the {Steam} platform.
\newblock {\em Procedia Computer Science}, 126:1289--1297, 2018.

\bibitem{becker2012analysis}
Roi Becker, Yifat Chernihov, Yuval Shavitt, and Noa Zilberman.
\newblock An analysis of the {Steam} community network evolution.
\newblock In {\em 2012 IEEE 27th Convention of Electrical and Electronics
  Engineers in Israel}, pages 1--5. IEEE, 2012.

\bibitem{drachen2016stylized}
Anders Drachen, Nicholas Ross, Julian Runge, and Rafet Sifa.
\newblock Stylized facts for mobile game analytics.
\newblock In {\em 2016 IEEE conference on computational intelligence and games
  (CIG)}, pages 1--8. IEEE, 2016.

\bibitem{engelstatter2018strategic}
Benjamin Engelst{\"a}tter and Michael~R Ward.
\newblock Strategic timing of entry: evidence from video games.
\newblock {\em Journal of Cultural Economics}, 42(1):1--22, 2018.

\bibitem{foxman2020virtual}
Maxwell Foxman, Alex~P Leith, David Beyea, Brian Klebig, Vivian Hsueh~Hua Chen,
  and Rabindra Ratan.
\newblock Virtual reality genres: Comparing preferences in immersive
  experiences and games.
\newblock In {\em Extended Abstracts of the 2020 Annual Symposium on
  Computer-Human Interaction in Play}, pages 237--241, 2020.

\bibitem{grewal2022empirical}
Balreet Grewal, Dayi Lin, Lars Doucet, and Cor-Paul Bezemer.
\newblock An empirical study of delayed games on {Steam}.
\newblock {\em arXiv preprint arXiv:2204.11191}, 2022.

\bibitem{heintz2015game}
Stephanie Heintz and Effie Lai-Chong Law.
\newblock The game genre map: A revised game classification.
\newblock In {\em Proceedings of the 2015 Annual Symposium on Computer-Human
  Interaction in Play}, pages 175--184, 2015.

\bibitem{li2020towards}
Xiaozhou Li.
\newblock Towards factor-oriented understanding of video game genres using
  exploratory factor analysis on {Steam} game tags.
\newblock In {\em 2020 IEEE International Conference on Progress in Informatics
  and Computing (PIC)}, pages 207--213. IEEE, 2020.

\bibitem{li2020preliminary}
Xiaozhou Li and Boyang Zhang.
\newblock A preliminary network analysis on {Steam} game tags: another way of
  understanding game genres.
\newblock In {\em Proceedings of the 23rd International Conference on Academic
  Mindtrek}, pages 65--73, 2020.

\bibitem{lieu2021trailer}
Derek Lieu.
\newblock Game trailer outline based on genre, 2021.
\newblock URL:
  \url{https://www.derek-lieu.com/blog/2021/2/28/game-trailer-outline-based-on-genre}.

\bibitem{lieu2021game}
Derek Lieu.
\newblock Game trailer structure - genre, hook, content, 2021.
\newblock URL:
  \url{https://www.derek-lieu.com/blog/2021/4/12/game-trailer-structure-genre-hook-content?utm_source=substack&utm_medium=email}.

\bibitem{lin2018empirical}
Dayi Lin, Cor-Paul Bezemer, and Ahmed~E Hassan.
\newblock An empirical study of early access games on the {Steam} platform.
\newblock {\em Empirical Software Engineering}, 23(2):771--799, 2018.

\bibitem{lucas2004sex}
Kristen Lucas and John~L Sherry.
\newblock Sex differences in video game play: A communication-based
  explanation.
\newblock {\em Communication research}, 31(5):499--523, 2004.

\bibitem{o2016condensing}
Mark O'Neill, Elham Vaziripour, Justin Wu, and Daniel Zappala.
\newblock Condensing steam: Distilling the diversity of gamer behavior.
\newblock In {\em Proceedings of the 2016 internet measurement conference},
  pages 81--95, 2016.

\bibitem{orland2014introducing}
Kyle Orland.
\newblock {Steam} tags, 2014.
\newblock URL:
  \url{https://arstechnica.com/gaming/2014/04/steam-gauge-addressing-your-questions-and-concerns/}.

\bibitem{petrosino2022panorama}
Simone Petrosino, Enrica Loria, Alexander Kainz, and Johanna Pirker.
\newblock The panorama of {Steam} multiplayer games (2018-2020): A player
  reviews analysis.
\newblock In {\em FDG'22: Proceedings of the 17th International Conference on
  the Foundations of Digital Games}, pages 1--7, 2022.

\bibitem{pinelle2008using}
David Pinelle, Nelson Wong, and Tadeusz Stach.
\newblock Using genres to customize usability evaluations of video games.
\newblock In {\em Proceedings of the 2008 conference on future play: Research,
  play, share}, pages 129--136, 2008.

\bibitem{pirker2022virtual}
Johanna Pirker, Enrica Loria, Alexander Kainz, Johannes Kopf, and Andreas
  Dengel.
\newblock Virtual reality and education--the {Steam} panorama.
\newblock In {\em FDG'22: Proceedings of the 17th International Conference on
  the Foundations of Digital Games}, pages 1--11, 2022.

\bibitem{qaffas2020operational}
Alaa Qaffas.
\newblock An operational study of video games’ genres.
\newblock 2020.

\bibitem{ratan2021gender}
Rabindra Ratan, Vivian Hsueh~Hua Chen, Frederik De~Grove, Johannes Breuer,
  Thorsten Quandt, and J~Patrick Williams.
\newblock Gender, gaming motives, and genre: Comparing singaporean, german, and
  american players.
\newblock {\em IEEE Transactions on Games}, 14(3):456--465, 2021.

\bibitem{samarasinghe2021data}
Dilini Samarasinghe, Michael Barlow, Erandi Lakshika, Timothy Lynar, Nour
  Moustafa, Thomas Townsend, and Benjamin Turnbull.
\newblock A data driven review of board game design and interactions of their
  mechanics.
\newblock {\em IEEE Access}, 9:114051--114069, 2021.

\bibitem{sifa2015large}
Rafet Sifa, Anders Drachen, and Christian Bauckhage.
\newblock Large-scale cross-game player behavior analysis on {Steam}.
\newblock In {\em Eleventh Artificial Intelligence and Interactive Digital
  Entertainment Conference}, 2015.

\bibitem{sobkowicz2016steam}
Antoni Sobkowicz and Wojciech Stokowiec.
\newblock {Steam} review dataset-new, large scale sentiment dataset.
\newblock In {\em Proceedings of the Tenth International Conference on Language
  Resources and Evaluation (LREC 2016) Workshop Emotion and Sentiment
  Analysis}, pages 55--58, 2016.

\bibitem{steamworks2021steam}
Steamworks.
\newblock {Steam} tags, 2021.
\newblock URL: \url{https://partner.steamgames.com/doc/store/tags#13}.

\bibitem{su2021comprehensive}
Yanhui Su, Per Backlund, and Henrik Engstr{\"o}m.
\newblock Comprehensive review and classification of game analytics.
\newblock {\em Service Oriented Computing and Applications}, 15(2):141--156,
  2021.

\bibitem{tekofsky2016effect}
Shoshannah Tekofsky, Paul Miller, Pieter Spronck, and Kevin Slavin.
\newblock The effect of gender, native english speaking, and age on game genre
  preference and gaming motivations.
\newblock In {\em International Conference on Intelligent Technologies for
  Interactive Entertainment}, pages 178--183. Springer, 2016.

\bibitem{vermeulen2016play}
Lotte Vermeulen and Jan Van~Looy.
\newblock “{I} play so {I} am?” a gender study into stereotype perception
  and genre choice of digital game players.
\newblock {\em Journal of Broadcasting \& Electronic Media}, 60(2):286--304,
  2016.

\bibitem{voorhees2014routledge}
Gerald Voorhees.
\newblock Chapter 31: Shooting.
\newblock In Mark~JP Wolf and Bernard Perron, editors, {\em The Routledge
  companion to video game studies}, chapter~31, pages 251--258. 2014.

\bibitem{windleharth2016full}
Travis~W Windleharth, Jacob Jett, Marc Schmalz, and Jin~Ha Lee.
\newblock Full {Steam} ahead: A conceptual analysis of user-supplied tags on
  {Steam}.
\newblock {\em Cataloging \& Classification Quarterly}, 54(7):418--441, 2016.

\bibitem{zuo2018sentiment}
Zhen Zuo.
\newblock Sentiment analysis of {steam} review datasets using naive bayes and
  decision tree classifier.
\newblock 2018.

\end{thebibliography}

\newpage
\appendix

\section{List of video game genres}\label{app:list_genres}

\begin{itemize}
    \item 2D Fighter, 2D Platformer, 3D Fighter, 3D Platformer, 4X
    \item Action, Action RPG, Action Roguelike, Action-Adventure, Adventure, Arcade, Archery, Arena Shooter, Auto Battler, Automation, Automobile Sim
    \item Base-Building, Battle Royale, Beat 'em up, Board Game, Boxing, Building, Bullet Hell
    \item CRPG, Card Battler, Card Game, Casual, Character Action Game, Chess, Choose Your Own Adventure, City Builder, Clicker, Collectathon, Colony Sim, Combat Racing, Cooking, Creature Collector
    \item Dating Sim, Deckbuilding, Diplomacy, Dungeon Crawler
    \item Education, Exploration
    \item Farming Sim, Fighting, Fishing, Flight, FPS
    \item God Game, Grand Strategy
    \item Hack and Slash, Heist, Hero Shooter, Hidden Object, Horror, Hunting
    \item Idler, Immersive Sim, Interactive Fiction
    \item JRPG
    \item Life Sim, Looter Shooter
    \item Management, Match 3, Metroidvania, Mining, MMORPG, MOBA, Mystery Dungeon
    \item Naval Combat
    \item On-Rails Shooter, Open World Survival Craft, Otome
    \item Parkour, Party Game, Party-Based RPG, Platformer, Point \& Click, Political Sim, Precision Platformer, Programming, Puzzle, Puzzle-Platformer
    \item RPG, Racing, Real Time Tactics, Rhythm, Rogue-like, Rogue-lite, Roguelike Deckbuilder, RTS, Runner
    \item Sandbox, Shoot 'Em Up, Shooter, Side Scroller, Simulation, Sokoban, Solitaire, Souls-like, Space Sim, Spectacle fighter, Sports, Stealth, Strategy, Strategy RPG, Survival, Survival Horror
    \item Tabletop, Tactical RPG, Third-Person Shooter, Time Management, Top-Down Shooter, Tower Defense, Trading, Trading Card Game, Traditional Roguelike, Trivia, Turn-Based Strategy, Turn-Based Tactics, Twin Stick Shooter, Typing
    \item Vehicular Combat, Visual Novel
    \item Walking Simulator, Wargame, Word Game

\end{itemize}

\end{document}